\documentclass[final,5p,twocolumn]{elsarticle}

\usepackage{graphicx}
\usepackage{dcolumn}
\usepackage{bm}
\usepackage{color}
\usepackage{amssymb}
\usepackage{epstopdf}
\usepackage{amstext}
\begin{document}

\begin{frontmatter}

\title{Why the measured cosmological constant is small}
\author{T. Rostami$^{1}$} \ead{t\_rostami@sbu.ac.ir}
\author{S. Jalalzadeh$^{2}$ }\ead{shahram.jalalzadeh@unila.edu.br}
\address{ $^{1}$Department of Physics, Shahid Beheshti University, Evin, Tehran 19839 Iran}
\address{ $^{2}$Federal University of Latin-America Integration, Technological Park of Itaipu, PO box 2123, Foz do Igua\c{c}u-PR,  85867-670, Brazil}
\date{\today}
\begin{abstract}
In a quest to explain the small value of the today's cosmological constant, following the approach introduced in \cite{Jalalzadeh:2013wza}, we show that the theoretical value of cosmological constant is consistent with its observational value. In more detail, we study the Freidmann-Lama\^{\i}tre-Robertson-Walker  cosmology  embedded isometrically in an $11$-dimensional ambient space.
The field equations determines $\Lambda$ in terms of other measurable fundamental
constants.  Specifically, it predicts that   the cosmological constant measured today
be $\Lambda L^2_{\text{Pl}}=2.56\times10^{-122}$,  as observed.
\end{abstract}
\end{frontmatter}

\section{Introduction}
\label{1}
One of the most controversial problems in physics is the cosmological constant problem (CCP).
The natural value predicted for the cosmological constant by particle physics is $\rho_{\Lambda}\sim M_{\text{Pl}}^{4}\sim (10^{18} GeV )^{4}$,
where $M_{\text{Pl}}$ is the Planck scale, which has a great discrepancy with the observational bound,
$\rho_{\Lambda}\sim (10^{-3} eV)^{4}$ (about 120 orders of magnitude) \cite{Ade:2013zuv}.
The supersymmetric theories put a lower bound for the cosmological constant which is about $\rho_{\Lambda}\sim M_{\text{SUSY}}^{4}$,
with $M_{\text{SUSY}}\gtrsim 1\, TeV$, the supersymmetry breaking scale,
yet has a difference of about $60$ orders of magnitude. This discrepancy between the expected
and observed value of cosmological constant remains as an unanswered question \cite{1}.
Many attempts have been done to give an acceptable and relevant answer to this problem in the context of general relativity, while some authors were seeking through the anthropic principle \cite{ant}. A different aperture arose when it was shown that the cosmological constant became zero as a result of a unbroken supersymmetry \cite{Zumino:1974bg}. Notwithstanding, many attempts done in the context of supergravity \cite{sg} and superstring \cite{ss}, yet there is no mechanism  known to have an effective cosmological constant which is of the same order of the observational limit and particularly is unaffected from the quantum fluctuation.
On the other hand, the discovery of the cosmic acceleration, directed the study of cosmological constant in another point of view. The positive cosmological constant was demanded by observations including the more detailed studies of supernovae \cite{2}, large-scale structure, and the cosmic microwave background radiation \cite{3}. Recently, the released Planck data combined with other astrophysical data, including Type Ia supernovae, constrained the equation of state of dark energy to $w = -1.006\pm 0.045$, which is well consistent with the expected value for a cosmological constant \cite{Planck:2015xua}.

Meanwhile, the braneworld scenarios present a solution to the unsolved problems in gravity and cosmology, such as dark energy. In addition, it brought new methods for having a vanishing cosmological constant on the brane \cite{path,Ru}, but non could provide a convincing and robust justification. So there is not a general solution to explain why it is non-zero and yet it is so small.
In the herein paper, we focus our attempts on the approach introduced in \cite{Jalalzadeh:2013wza} and try to address this problem. In that paper, a covariant or a model independent Einstein's equations of a brane world model embedded isometrically and locally  in a ambient space with arbitrary number of dimensions are derived.
Therein, the use of Nash's theorem for the perturbation of the submanifold lets the brane to have a thickness. In brane models with one non-compact extra dimension, the extrinsic curvature is defined by using the junction condition. But, in the case of multiple extra dimensions, the junction condition is not applicable. Then additional assumptions are needed to specify the extrinsic curvature.
It was shown that, the induced gravitational constant, $G_{N}$, depended on the local normal radii of the  Freidmann-Lama\^{\i}tre-Robertson-Walker (FLRW) submanifold, an arbitrary function of the cosmic time \cite{Jalalzadeh:2013wza}.
In this paper, to be able to solve the resulting field equations, we will use a simple phenomenological power law form of $G_N$ in terms of scale factor.
Consequently the induced extrinsic term on the field equations of brane is of cosmological constant type.
We will show that the observed cosmological constant is a gravitational-geometrical constant, and it is not related to the vacuum energy.

\section{Extrinsic gravity}
Following our previous work \cite{Jalalzadeh:2013wza},  consider the $D$-dimensional
Minkowskian ambient space $(\mathcal M_D,\eta)$ with local natural Cartesian
coordinates $\mathcal Y^A=\{\mathcal Y^0,...,\mathcal Y^{D-1}\}$, endowed with a metric $\eta$ with signature $(-,+,...,+)$. Furthermore, consider in $\mathcal M_D$ a $4D$ Lorentzian submanifold $(\mathcal M_4,g)$ with adapted
local coordinates $x^\mu = \{x^0, . . . , x^3\}$ and induced metric $g$. We can then construct
an adopted coordinate system in the ambient space which includes the local coordinates
of submanifold $\mathcal M_4$ as $\{x^\mu, x^a\}$, where $x^a = \{x^4, . . . , x^{D-1}\}$ are extrinsic or
extra coordinates. In such a condition, the submanifold $\mathcal M_4$ is defined by $x^a = 0$. Therefore, the isometric local embedding is given by $D$ differential maps
\begin{eqnarray}\label{sh1}
\mathcal Y^A: \mathcal M_4\rightarrow \mathcal M_D.
\end{eqnarray}
Also, the vectors
\begin{eqnarray}\label{sh2}
e_\mu=\mathcal Y^A_{,\mu}\partial_A,\,\,\,e_a=\mathcal N^A_a\partial_A,
\end{eqnarray}
form a basis of tangent and normal vector spaces respectively at each point of $\mathcal M_4$, where $\mathcal N^A_a$ denotes the components
of $n = D-4$ unit vector fields orthogonal to the $\mathcal M_4$ and also normal to each other
in direction of the extra coordinates $x^a$. Consequently, differential map $\mathcal Y^A$ satisfies embedding equations \footnote{Greek indices run from 0 to 3, small case Latin indices
 run from 4 to $D-1$ and large Latin indices run from 0 to $D-1$. Units  so
that $\hbar=c=1$ are used throughout this work. Also, $\partial_\mu:=\frac{\partial}{\partial
x^\mu}$, $\partial_A:=\frac{\partial}{\partial{\mathcal Y}^A}$ and ${\mathcal
Y}^A_{,\mu}:=\partial_\mu\mathcal Y^A$.}
\begin{eqnarray}\label{sh3}
\begin{array}{cc}
\eta(e_\mu,e_\nu)=g_{\mu\nu}(x^\alpha),\\
\eta(e_\mu,e_a)=0,\\
\eta(e_a,e_b)=\delta_{ab}.
\end{array}
\end{eqnarray}
Then the projected gradients of bases  vectors can be decomposed with
respect to the bases vectors $\{e_\mu,e_a\}$ which are the generalizations of well-known
Gauss-Weingarten equations
\begin{eqnarray}\label{sh4}
\begin{array}{cc}
e_{\mu,\nu}=\Gamma^\alpha_{\mu\nu}e_\alpha+K_{\mu\nu}^ae_a,\\
e_{a,\mu}=-K_{\mu\nu a}e^\nu+A_{\mu a}^{\,\,\,b}e_b,
\end{array}
\end{eqnarray}
where $\Gamma^\alpha_{\mu\nu}$ are the connection coefficients compatible with induced metric, $K_{\mu\nu a}:=-\eta(e_{a,\mu},e_\nu)$ denotes extrinsic curvature and $A_{\mu ab}$ is the extrinsic twist potential (or third fundamental form) defined by $A_{\mu ab}:=\eta(e_{a,\mu},e_b)=-A_{\mu ba}$. To introduce a brane with constant thickness $l$, according to the Nash-Morse implicit function theorem \cite{Nash}, the submanifold $(\mathcal M_4,g)$ is deformed so that it remains compatible with confinement. With this method one can generate a sequences of $4D$ submanifolds isometrically embedded in ambient space, inside of thickness. Nash's strategy, which is also
the strategy of later authors, was to consider the problem of perturbing
a given isometric  embedding to achieve some desired and suitably small change in the metric \cite{Ham, Maia:1984nv,Maia:2011yu}. Suppose the local coordinates of perturbed submanifold in the vicinity of original one, $(\mathcal M_4,g)$, is given by
\begin{eqnarray}\label{sh5}
\mathcal Z^A(x^\mu,x^a)=\mathcal Y^A(x^\mu)+\sqrt{\sigma}x^a\mathcal N^A,
\end{eqnarray}
where $\sigma$ denotes small perturbation parameter.
Then, the deformed or perturbed embedding will be \cite{Jalalzadeh:2013wza}
\begin{eqnarray}\label{sh6}
\begin{array}{cc}
\mathcal Z^A_{,\mu}=\mathcal Y^A_{,\alpha}(\delta^\alpha_\mu-\sqrt{\sigma}x^aK^{a\,\,\,\,\alpha}_{\,\,\,\,\mu})+\sqrt{\sigma}x^mA_{\mu m}^{\,\,\,\,\,\,n}\mathcal N^A_n,\\
\mathcal Z^A_{,a}=\sqrt{\sigma}\mathcal N^A_a.
\end{array}
\end{eqnarray}
The line element of ambient
space in the vicinity of submanifold is $ds^2=\eta_{AB}d\mathcal Z^Ad\mathcal Z^B$. By substituting perturbed embedding (\ref{sh6}) into the expression for the line element, the metric of ambient space in the Gaussian coordinates $\{x^\mu,x^a\}$ will be
\begin{eqnarray}\label{sh7}
\eta_{AB}=\left(
  \begin{array}{cc}
    \gamma_{\mu\nu}+\sigma A_{\mu a}A_\nu^{\,\,a} & \sigma A_{\mu m} \\
    \sigma A_{\nu n} & \sigma\delta_{mn} \\
  \end{array}
\right),
\end{eqnarray}
where $A_{\mu m}=x^nA_{\mu nm}$ and
\begin{eqnarray}\label{sh8}
\begin{array}{cc}
\gamma_{\mu\nu}=\\
g^{\alpha\beta}\left(g_{\mu\alpha}-\sqrt{\sigma}x^mK_{\mu\alpha m}\right)\left(g_{\nu\beta}-\sqrt{\sigma}x^nK_{\nu\beta n}\right).
\end{array}
\end{eqnarray}
This relation guides  us to the definition of the curvature radii of the submanifold $(\mathcal M_4,g)$: The extreme values of normal curvature are calculated by the homogeneous equations
\begin{eqnarray}\label{sh9}
\left(g_{\mu\nu}-L^a_{(\mu)}K_{\mu\nu a}\right)\delta x^\mu=0,
\end{eqnarray}
where  $L^a_{(\mu)}$ are the curvature radii for each principal direction
$\delta x^\mu$ and for each normal $e_a$.
 Hence the normal curvature determines the shape of the submanifold in the neighbourhood of any points. Then the extreme value of the normal curvature is obtained by
\begin{eqnarray}\label{1-18}
\det\left({g}_{\mu\alpha}-L^a_{(\mu)}{K}_{\mu\alpha
a}\right)=0.
\end{eqnarray}
It follows that the metric of deformed submanifold (\ref{sh8}) becomes singular at the solution of equation (\ref{1-18}).
Hence, at each point of submanifold, the normal curvature radii generates a closed space ${\mathcal B}_{n}$ (Extrinsic tube). The physical space on the ambient space is bounded to the Extrinsic tube as shown in Fig. (1). All extra dimensions
are assumed to be spacelike, then ${\mathcal B}_{n}$ may be taken locally to be the $n$-sphere
$S_{n}=SO(n)/SO(n-1),$
with radius  $L:=min\{L^a_{(\mu)}\}$, at each point of spacetime \cite{Maia:1984nv}.
According to the confinement hypothesis, it is supposed that the standard model particles are localized to a $4$-dimensional submanifold, while gravity can freely propagate in the ambient space. However, due to equation (\ref{1-18}), the normal curvature radii determines a characteristic radius, for the propagation of the gravitons.

Generally, by substitution of ${\mathcal Z}^A_{,\mu}$ derived from (\ref{sh6})
into the  extrinsic curvature of deformed submanifold, defined by $K_{\mu\nu
a}(x^\alpha,x^b):=-\delta_{AB}{\mathcal N}^A_{;\mu}{\mathcal Z}^B_{,\nu}$, we
obtain the extrinsic curvature of deformed submanifold in term of extrinsic
curvature of original non-perturbed submanifold as
\begin{eqnarray}\label{sha1}
\begin{array}{cc}
K_{\mu\nu a}(x^\alpha,x^b)=\\
K_{\mu\nu a}(x^\alpha)-\frac{\sqrt{\sigma}}{2}x^m\left(K_{m\gamma(\mu }K_{\nu)\,\,\,\,a}^{\,\,\,\,\gamma}-F_{\mu\nu an}\right).
\end{array}
\end{eqnarray}
Comparing definition of $\gamma_{\mu\nu}$ in (\ref{sh8}) with (\ref{sha1})
we obtain
\begin{eqnarray}\label{sha2}
K_{\mu\nu a}(x^\alpha,x^b)=-\frac{1}{2\sqrt{\sigma}}\partial_a\gamma_{\mu\nu}+\frac{\sqrt{\sigma}}{2}x^bF_{\mu\nu
ab},
\end{eqnarray}
where $F_{ab\mu\nu}$ is the curvature associated with extrinsic twist vector field $A_{\mu ab}$, defined as  \cite{twist}
\begin{eqnarray}
\label{1-10}
F_{ab\mu\nu}=A_{\mu ab,\nu}-A_{\nu ab,\mu}-A_{\nu a}^{\,\,\,\,\,\,\,c}A_{\mu
cb}+A_{\mu a}^{\,\,\,\,\,\,\,c}A_{\nu cb},
\end{eqnarray}
where $A_{\mu ab}$ plays the role of the Yang-Mills potential \cite{Yang}. Notice that $A_{\mu ab}$ transform as the component of a gauge vector field under the group of isometries of the bulk if the bulk space (ambient space) has certain
Killing vector fields \cite{Shahram} and it only exist in the ambient spaces with dimensions equal to or
greater than six $(n\geqslant 2)$.
The extrinsic curvature gives a measure
of the deviation from the submanifold and its tangent plane at any point.
Its symmetric part shows that the second fundamental form
propagates in the bulk. The antisymmetric part is proportional to a Yang-Mills
gauge field, which really can be thought of as a kind of curvature.

\begin{figure}
  \centering
  \includegraphics[width=8cm]{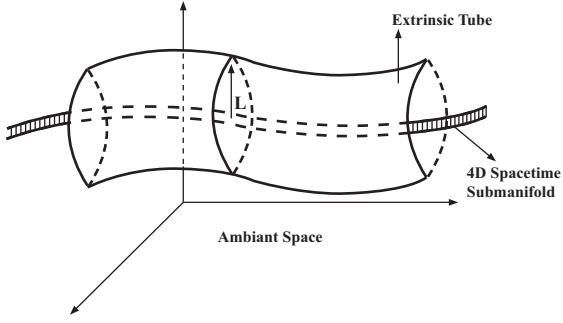}\\
  \caption{\small  Brane with thickness $l$ is embedded in $D$-dimensional ambient space which is bounded by a closed space (Extrinsic tube), generated by the normal curvature radii $L$. }\label{fig}
\end{figure}
Now, using expression (\ref{sh8}) in Einstein-Hilbert action functional of abient spacetime and remembering the confinement
hypothesis (matter and gauge fields are confined to the brane with thickness $l$
while gravity could propagate in the ambient space up to the curvature radii $L$ with
the local geometry of $S_n$) we obtain
\begin{eqnarray}\label{1-30004}
\begin{array}{cc}
-\frac{1}{2\kappa^2_D}\int{\sqrt{|\mathcal{G}|}{\cal R}d^Dx}\simeq\\
\\
-\frac{M_D^{n+2}\sigma^{n/2}V_n}{16\pi}\int
L^n\sqrt{|\bar{g}|}\left(\bar{R}-\bar{K}_{\alpha\beta m}{\bar K}^{\alpha\beta m}
+ \bar{K}^2\right)d^4x\\
\\
+\frac{nM_D^{n+2}\sigma^{n/2}l^3V_n}{64\pi(n+2)}\int L^{n-1}\sqrt{|\bar{g}|}F_{\mu\nu
mn}F^{\mu\nu mn}d^4x,
\end{array}
\end{eqnarray}
where $V_n=\pi^{\frac{n}{2}}/\Gamma(\frac{n}{2}+1)$. Hence, the relation between the fundamental scale $M_D$ and the $4D$ Planck
scale $M_{Pl}$ will be
\begin{eqnarray}\label{sh11}
M^2_{\text{Pl}}=\frac{\pi^\frac{n}{2}}{\Gamma(\frac{n}{2}+1)}L^n\sigma^\frac{n}{2}M_{D}^{n+2}.
\end{eqnarray}
Also, with an eye on the Lagrangian density of the Yang-Mills field
\cite{Scheck}
\begin{eqnarray}\label{f1}
{\mathcal L}^{(YM)}=\frac{1}{4g_i^2\kappa'}tr(F_{\mu\nu}F^{\mu\nu}),
\end{eqnarray}
the last part of (\ref{1-30004}) gives
\begin{eqnarray}\label{sh12}
\frac{4\pi}{\kappa'g_i^2}=\frac{n\pi^\frac{n}{2}M_D^{n+2}\sigma^\frac{n}{2}l^3L^{n-1}}{4(n+2)\Gamma(\frac{n}{2}+1)},
\end{eqnarray}
where $g_i$ are the gauge couplings and the index $i$ labels the simple subgroups of
the gauge group with
\begin{eqnarray}\label{sh13}
\kappa'=\left\{
                           \begin{array}{ll}
                             2,\,\,\,\,\,\,\,\,\,\,\,\,\,\,\,\,\,\,\,\,\,n=2, \\
                             2(n-2),\,\,\,n>2.
                           \end{array}
                         \right.
\end{eqnarray}
Therefore, the induced action functional is equivalent to the $4D$ gravitational action containing extrinsic terms and Yang-Mills action
\begin{eqnarray}\label{sh10}
\begin{array}{cc}
-\frac{1}{2\kappa_D}\int \sqrt{|\eta|}\mathcal Rd^Dx=\\
-\int \frac{M^2_{Pl}}{2}\sqrt{|g|}\left(R-K_{\alpha\beta a}K^{\alpha\beta a}+K_aK^a\right)d^4x+\\+\int\frac{1}{4 \kappa'g_i^2}tr (F_{\mu\nu}F^{\mu\nu})d^4x,
\end{array}
\end{eqnarray}
Equations (\ref{sh11})
and (\ref{sh12}) immediately give us the following fundamental relation between normal
curvature radii, thickness of the brane, number of extra dimensions and 4D Planck's mass
\begin{eqnarray}\label{1,35,0}
L=l^3M_{\text{Pl}}^{2}\kappa\frac{g_{i}^{2}}{4\pi},
\end{eqnarray}
in which $\kappa$ is given by
\begin{eqnarray}\label{1-35}
\kappa=\left\{
                      \begin{array}{ll}
                        \frac{n}{2(n+2)}, & \hbox{$n=2$,} \\
                       \frac{n(n-2)}{2(n+2)}, & \hbox{$n>2$.}
                      \end{array}
                    \right.
\end{eqnarray}
Note that if $L\sim l\sim L_{\text{Pl}}$, the above equation reduces to the equivalent relation
in Kaluza-Klein gravity \cite{Love}. Also, from equation (\ref{1,35,0}),
it is easy to see that the following relation is hold between gravitational
``constant'' and gauge couplings
\begin{eqnarray}\label{r1}
\frac{G_N}{G_0}=\left(\frac{g_i^2}{g^2_{0i}}\right)^{-\frac{n}{n+1}},
\end{eqnarray}
where $G_0$ and $g_{0i}$ are the gravitational constant and gauge coupling constants
at the present epoch, respectively.

Then, the induced Einstein-Yang-Mills field equations for the thick brane world model will be \cite{Jalalzadeh:2013wza}
\begin{eqnarray}
\label{1-38}
\begin{array}{lll}
{G}_{\alpha\beta}= -Q_{\alpha\beta}+8\pi G_N\left(T_{\alpha\beta}+T_{\alpha\beta}^{(YM)}\right),\\
\\
\nabla^{(tot)}_\beta{K}_a-\nabla^{(tot)}_\alpha{K}_{\beta a}^{\alpha}=8\pi
G_NT_{a\beta} ,\\
\\
\frac{G_N}{\kappa' g_i^2}\left(F^{\alpha\beta}_{\,\,\,\,\,\,\,am}F_{\alpha\beta
b}^{\,\,\,\,\,\,\,\,\,\,m}+\frac{1}{2}\eta_{ab}F_{\alpha\beta}^{\hspace{.3cm}lm}F^{\alpha\beta}_{\hspace{.3cm}lm}\right)-
\\
-\frac{1}{2}\eta_{ab}\left({R}+{K}_{\mu\nu m}{K}_{\mu\nu}^{\hspace{.3cm}m}-{K}_a{K}^a\right)=8\pi
G_NT_{ab},
\end{array}
\end{eqnarray}
where ${G}_{\alpha\beta}$ is the $4D$ Einstein tensor and  $G_{N}$ is the induced gravitational constant.
$T_{\alpha\beta}$, $T_{a\beta}$ and $T_{ab}$ are the components of energy-momentum tensor defined such that to be compatible with the confinement hypothesis.
$T_{\alpha\beta}^{(YM)}$ denotes the Yang-Mills energy-momentum tensor and $Q_{\alpha\beta}$ is a conserved quantity
expressed in terms of extrinsic curvature and its  trace, $K_a$, as
\begin{eqnarray}\label{1-0}
\begin{array}{lll}
Q_{\alpha\beta}={K}_{\alpha}^{\,\,\,\,\eta a}{K}_{\beta\eta a}-{K}^a{K}_{\alpha\beta a}-\\ -\frac{1}{2}{g}_{\alpha\beta}({K}^{\mu\nu
a}{K}_{\mu\nu a}-{K}_a{K}^a),\\
\\
Q^{\alpha\beta}_{\,\,\,\,\,\,\,\,;\beta}=0,
\end{array}
\end{eqnarray}
and $\nabla_\mu^{(tot)}$, the total covariant derivative, is defined as
\begin{eqnarray}\label{1-27}
\nabla_\mu^{(tot)}{K}_{\alpha\beta m}={K}_{\alpha\beta m;\mu}-A_{\mu
mn}{K}_{\alpha\beta}^{\,\,\,\,\,\,n}.
\end{eqnarray}
The resulting field equations provide both the equations of general relativity and of Yang-Mills. Thus, it introduces a unifying picture like the Kaluza-Klein theory \cite{Kaluza:1921tu}.

\section{FLRW cosmology}
Let us now analyze the influence of the extrinsic curvature
terms on a Homogeneous  and isotropic universe. We assume that the FLRW spacetime is embedded locally and isometrically in a $D=n+4$ dimensional Minkowskian spacetime. For simplicity, we assume the twisting vector fields $A_{\mu ab}$ vanish (charge less universe). Consider the standard spatially homogeneous and isotropic  line element
\begin{eqnarray}\label{2-1}
ds^2=-dt^2+a(t)^2\left(\frac{dr^2}{1-kr^2}+r^2d\Omega^2\right),
\end{eqnarray}
where $a(t)$ is the cosmic scale factor and $k$ is $+1$, $-1$ or $0$, corresponding
to the closed, open, or flat universes, respectively.
 Due to the symmetries
of the embedded 4-dimensional universe, the energy-momentum tensor
is taken to be diagonal. Hence, we adopt the a perfect fluid form for the energy-momentum tensor in comoving
coordinates, as
\begin{eqnarray}\label{2-2}
\begin{array}{lll}
T_{\mu\nu}= (\rho+p)u_\mu u_\nu+pg_{\mu\nu},~~u_{\mu}=-\delta_{\mu}^{0},\\
\\
T_{\mu a}=0,\\
\\
T_{ab}=p_{ext.}\eta_{ab},
\end{array}
\end{eqnarray}
where $p_{ext.}$ is the pressure along the extra dimensions, inside of thick brane.

As we know, the line element (\ref{2-1}) is conformally flat. On the other
hand, it is well known that  a submanifold of a conformally flat ambient space is conformally flat if and only if it is totally quasiumbilical and a totally quasiumbilical submanifold of a conformally flat space is always conformally flat \cite{Chen}. This means that for a totally quasiumbilical submanifold, the second fundamental form is given by
\begin{eqnarray}\label{1,1}
{K}_{\alpha\beta m}=D_{m}g_{\alpha\beta}+B_{m}u_{\alpha}u_{\beta},
\end{eqnarray}
where $D_{m}$ and $B_{m}$ are arbitrary functions.
In particular, if $D_{m}=0$, then the submanifold is said to be cylindrical and if $B_{m}=0$, then the submanifold is said to be totally umbilical.
One can obtain the components of extrinsic curvature from  the second set of the field equations (\ref{1-38}) as
\begin{eqnarray}\label{2-3}
\begin{array}{lll}
K_{00m}=-\frac{1}{a(t)H}\frac{d}{dt}\left(\frac{f_m(t)}{a(t)}\right),\\
\\
{K}_{\alpha\beta m}=\frac{f_m(t)}{a^2(t)}g_{\alpha\beta},\hspace{.5cm}\alpha,\beta=1,2,3,
\end{array}
\end{eqnarray}
where $f_m(t)$ are arbitrary functions of cosmic time $t$, and $H$ is the
Hubble parameter.
Note that, $D_{m}$ and $B_{m}$ in equation (\ref{1,1}) can be verified in comparison with equation (\ref{2-3}) as
\begin{eqnarray}\label{2,4}
\begin{array}{cc}
D_{m}=\frac{f_m(t)}{a^2(t)},\\
B_{m}=\frac{f_m(t)}{a^2(t)}-\frac{1}{a(t)H}\frac{d}{dt}\left(\frac{f_m(t)}{a(t)}\right),
\end{array}
\end{eqnarray}
which shows that our calculations are consistent with above theorem on the quasiumbilical submanifolds.
The values of normal curvature of FLRW universe
can be obtained from equation (\ref{1-18}) as
\begin{eqnarray}\label{2-5}
\begin{array}{lll}
L_{(0)}^m=\frac{a^2(t)}{f_m(t)},\\
\\
L_{(\alpha)}^m=\frac{\frac{a^2(t)}{f_m(t)}}{-1+\frac{\dot f_m(t)}{f_m(t)H}}, \hspace{.3cm}\alpha=1,2,3.
\end{array}
\end{eqnarray}
Note that to determine uniquely the extrinsic curvature, the arbitrary functions $f_{m}$ must be determined. Notwithstanding the fact that we considered more than one extra dimension, the junction condition is not applicable. Hence, extra assumptions are needed to determine the extrinsic curvature.

First,
the symmetries of spacetime lets us assume the functions
$f_m(t)$ to be equal; $f_m(t)=f(t)$. If we set $\phi(t)=\phi_m(t)=a^2(t)/f^m(t)$ and assume $\frac{\dot \phi(t)}{\phi(t)H}>0$, the normal curvature radii will be
\begin{eqnarray}\label{2-6}
L=min(L_{(0)},L_{(\alpha)})=\phi(t).
\end{eqnarray}
Therefore, according to equations (\ref{sh11}), (\ref{r1}) and (\ref{2-6}) the induced $4D$ gravitational ``constant'' and the fine structure ``constant'',
$\alpha$, are not actually  true constants
and  dependent on the local normal radii of $4D$ submanifold as
\begin{eqnarray}\label{2-7}
G_N=G_0\left(\frac{L}{L_0}\right)^{-n},\hspace{.2cm}\alpha=\alpha_0\left(\frac{L}{L_0}\right)^{n+1}
\end{eqnarray}
where $L_0$ is the present value of $\phi(t)=L$ and $\alpha_0$
is the fine structure at the present epoch.
On the other hand, using equations (\ref{sh7}) and (\ref{sh8}), the line element of ambient
space will be
\begin{eqnarray}\label{sha4}
\begin{array}{cc}
ds^2_{\text{ambient}}=-\left(1-\sqrt\sigma\frac{1}{\phi^m}x^m(1-\frac{h}{H})\right)^2dt^2+\\
\left(1-\frac{\sqrt{\sigma}}{\phi^m}x^m\right)^2g_{\mu\nu}(x^\alpha)dx^\mu
dx^\nu+\sigma\delta_{mn}dx^m dx^n,
\end{array}
\end{eqnarray}
where $h:=\dot L/L$ and $\mu,\nu=1,2,3$. 
Also, the extrinsic curvature of any perturbed submanifold, inside of thick
brane, using (\ref{sha1}) or (\ref{sha2})
will be
\begin{eqnarray}\label{sha6}
\begin{array}{cc}
K_{00a}(x^\alpha,x^m)=\\
=-\frac{1}{\phi^a}(1-\frac{h}{H})\left(1-\frac{\sqrt\sigma}{\phi^n}x^n(1-\frac{h}{H})\right),\\
K_{\mu\nu a}(x^\alpha,x^m)=\frac{1}{\phi^a}\left(1-\frac{\sqrt{\sigma}}{\phi^n}x^n\right)g_{\mu\nu}(x^\alpha),\\\mu,\nu=1,2,3.
\end{array}
\end{eqnarray}
Therefore, the deformed geometry will be well defined if we know two independent
fields: the metric
and the bending function  of non-perturbed
submanifold.

 The components of $Q_{\alpha\beta}$ defined in
(\ref{1-0}) for an quasiumbilical submanifold will be
\begin{eqnarray}\label{2-4}
\begin{array}{cc}
Q_{00}=\frac{3n}{L^2},\\
Q_{\mu\nu}=-\frac{3n}{L^2}\left(1-\frac{2h}{3H}\right)g_{\mu\nu},\hspace{.3cm}\mu,\nu=1,2,3.
\end{array}
\end{eqnarray}

Therefore, for FLRW universe, using equations (\ref{2-4}) in field equations (\ref{1-38}) plus generalized conservation equation for matter field $(G_NT^{\mu\nu})_{;\mu}=0$ for perfect fluid with equation
of state $p =\omega\rho$, the induced Friedmann equations will be
\begin{eqnarray}\label{2-9}
\begin{array}{lll}
H^2+\frac{k}{a^2}=\frac{8\pi G_{0}}{3}\rho_0\left(\frac{a}{a_0}\right)^{-3(1+\omega)}+\frac{n}{L^2},\\
\\
\frac{\ddot{a}}{a}=-\frac{4\pi G_{0}}{3}(1+3\omega)\rho_0\left(\frac{a}{a_0}\right)^{-3(1+\omega)}+\\
+\frac{n}{L^2}\left(1-\frac{h}{H}\right).
\end{array}
\end{eqnarray}
Also, determining the trace of the first equation in (\ref{1-38}) and plugging
 it into the third equation of (\ref{1-38}), leads to
\begin{eqnarray}\label{2-10}
3p-\rho=2p_{ext},
\end{eqnarray}
which shows that the constancy of thickness  gives a simple restriction on
the pressure component of the confined matter fields along the extra dimensions.
Since we do not yet Know the functional form of $L$ or $\phi(t)$, we have only the formal structure of a theory. To be able to solve above dynamical equations, we must specify $L$. For this purpose we use a
simple phenomenological form for gravitational  ``constant'' as
\begin{eqnarray}\label{r2}
G_N=G_0\left(\frac{a}{a_0}\right)^\delta,
\end{eqnarray}
which  using (\ref{2-7}) leads to
\begin{eqnarray}\label{r4}
L=L_0\left(\frac{a}{a_0}\right)^{-\frac{\delta}{n}},\hspace{.3cm}\alpha=\alpha_0\left(\frac{a}{a_0}\right)^{-\frac{\delta(n+1)}{n}}.
\end{eqnarray}
Consequently, the field equations  (\ref{2-9}) will be
\begin{eqnarray}\label{2-9aa}
\begin{array}{lll}
H^2+\frac{k}{a^2}=\frac{8\pi G_{0}}{3}\rho_0\left(\frac{a}{a_0}\right)^{-3(1+\omega)}+\frac{n}{L_0^2}\left(\frac{a}{a_0}\right)^{\frac{2\delta}{n}},\\
\\
\frac{\ddot{a}}{a}=-\frac{4\pi G_{0}}{3}(1+3\omega)\rho_0\left(\frac{a}{a_0}\right)^{-3(1+\omega)}+\\
+\frac{n}{L_0^2}\left(1+\frac{\delta}{n}\right)\left(\frac{a}{a_0}\right)^{\frac{2\delta}{n}}.
\end{array}
\end{eqnarray}
The behavior of $L$
near a spacetime singularity depends on the topological nature of that singularity.
In the point-like singularity (like FLRW case) all values of $L_{(\mu)}$
tend to zero so that $L$ also tends to zero. This property plus the form
of $L$ in equation (\ref{r4}) implies $\delta<0$. On the other hand, a recent fit of about 30 years of LLR data yields the excellent measurement
constraint on the time variation of $G_N$ \cite{GG}
\begin{eqnarray}\label{r5}
\frac{\dot G_N}{G_N}\simeq(0 \pm1.1)\times10^{-12}y^{-1},
\end{eqnarray}
which is a small fraction, about $\frac{1}{60}$, of the observed Hubble expansion rate of the universe. Hence, $\delta\simeq-\frac{1}{60}$.
As we see from the
 Friedmann equations (\ref{2-9aa}), the extrinsic curvature
plays the role of cosmological ``constant'', where
\begin{eqnarray}\label{2-25}
\begin{array}{lll}
\rho_{\Lambda}=\frac{3n}{8\pi G_{N}L^{2}}=\frac{3n}{8\pi G_{0}L_0^{2}}\left(\frac{a}{a_0}\right)^{\frac{1}{60}(1-\frac{2}{n})},\\
\\
p_{\Lambda}=(-1+\frac{1}{30n})\rho_{\Lambda}.
\end{array}
\end{eqnarray}
Therefore, the extrinsic shape of brane induces a cosmological ``constant''
proportional to the inverse-square of  normal curvature radii,
\begin{eqnarray}\label{r6}
\Lambda=8\pi G_N\rho_\Lambda=\Lambda_0\left(\frac{a}{a_0}\right)^{-\frac{1}{30n}},\hspace{.2cm}
\Lambda_0=\frac{3n}{L_0^2},
\end{eqnarray}
 Explicitly, by inserting equation (\ref{1,35,0}) into (\ref{r6}) we obtain
\begin{eqnarray}\label{2-26}
\begin{array}{cc}
\Lambda_0 L_{\text{Pl}}^2=\frac{3n}{\kappa^2}\left(\frac{4\pi}{g_0^2}\right)\left(\frac{L_{\text{Pl}}}{l}\right)^6,
\omega_\Lambda=-1+\frac{1}{30n},
\end{array}
\end{eqnarray}
where $L_\text{Pl}$ denotes reduced Planck's length at the present epoch.
An interesting point to note is that  the obtained cosmological constant is inversely proportional to the sixth power of brane thickness, the inverse
 of the coupling constants.

{  Let us now back to the number of extra dimensions in our model. According to the Janet-Cartan theorem \cite{Janet}, an Riemannian manifold of dimension $d$ can be embedded locally and isometrically into ambient Riemannian space of dimension $D$, if $D\geq\frac{d(d+1)}{2}$. Friedman generalized this theorem to the case of Lorentzian manifolds and showed that the same result is holed \cite{Fr}. Therefore, for a 4-dimensional embedded spacetime, the theorem ensures the existence of a local isometric embedding for $D\geq10$. On the other hand, one of the major goals of field equations (\ref{1-38}) is to find a geometrical method to unify fundamental forces, using the group of isometries of the ambient space \cite{Jalalzadeh:2013wza}.  When we identify
the extrinsic twist vector fields $A_{\mu ab}$  with the physical gauge fields, a simple arithmetic
fixes also the number of extra dimensions: For the standard model $U(1)\times SU(2)\times SU(3)$ it was found that the self-contained
gauge structure requires 11 dimensions \cite{Witten:1981me,Maraner:2007wk}.}

{ Let us return  to the thickness of brane.  The thickness of brane can be taken as a fundamental minimal length related to the
the trapped standard model interactions on the brane.  For example, let $l$ be of order of the Planck length, then according to equation  (\ref{2-26}) we have $\Lambda\sim M_{\text{Pl}}^2$ (or equivalently $\rho_{\Lambda}\sim M_{\text{Pl}}^4$). This is the value predicted by by particle physics field. At the opposite end, by considering $l\gg L_{\text{Pl}}$ it leads to a very small cosmological constant. In embedding models of gravity, we usually assume that the standard model matter fields are confined to the $4D$ submanifold. But in accordance with Heisenberg's principle the
confinement causes the matter fields to fluctuate very strongly in the direction
normal to the submanifold, along the extra dimensions. In other words, if we assume that $\Delta x^a=0$, then the momentum of confined particles along extra dimensions will be infinitely large. For example, trying to localize an particle to within less than its Compton wavelength makes its momentum so uncertain that it can have an energy large enough to make an extra particle-antiparticle pair. In this direction, there is another supportive  argument. Suppose
a massless Dirac (Klein-Gordon) equation in the ambient space. For a $D$-dimensional
spinorial field (scalar field) we may make the harmonic expansion on the
compact manifold constructed from thickness of brane \cite{Love}. Then the
induced fields are mass eigenstates in $4D$ spacetime, with $m_n\approx \frac{n}{l},$
where $n=0,1,2,...$ , which shows that the thickness is proportional with
the inverse of lowest mass, $n=1$, particles.
  Therefore, according to the quantum mechanics  the thickness of brane is proportional with the Compton wavelength of   trapped baryonic matter (since most of the mass of ordinary matter comes from the protons and neutrons inside atomic nuclei, and protons and neutrons are classified by particle physicists as baryons)  which can be thought of as a fundamental limitation on measuring the position of a particles.   In
the strong interactions physics, it is sometimes assumed that coupling to the lowest
mass mesons determines the localization size of hadrons to be $1/{m_\text{meson}}$ in the
manner first described by Yukawa \cite{Yokawa}.}
On the other hand, as we know, all hadrons couple via the strong interaction to low
mass mesons. Consequently, the thickness  of brane
cannot be appreciably less than the range of the strong interactions
$R_{had}\sim1/\lambda_{\text{QCD}}\sim 1$ fm.
Further evidence in support of the idea that all hadrons, e.g. nucleons, couple strongly
to low mass mesons is the fact that pions are the principal secondary component
in high-energy collisions \cite{Bushin}.
Therefore, in a good approximation $l$ is set to be the overlap of the quark and antiquark wave functions in the pions, $f_{\pi^{-}}=130.41$ $MeV$ \cite{Rosner:2012np} (size of all hadrons are $\gtrsim \frac{1}{f_{\pi^{-}}}$). Therefore, we adopt
\begin{eqnarray}\label{13-0}
l=\frac{1}{f_{\pi}}.
\end{eqnarray}
Inserting this relation into the equation (\ref{2-26}) and also $n=7$, we
obtain
\begin{eqnarray}\label{13}
\begin{array}{cc}
\rho_{0\Lambda}=
\frac{3}{2\pi}\left(\frac{4\pi}{g_0^2}\right)^2G_0 f_{\pi}^{6},\\
\omega_\Lambda=-0.998
\end{array}
\end{eqnarray}
where $g_0^2/4\pi=16.8$ is the strong interaction pion-nucleon coupling parameter
at the present epoch
\cite{Ioffe:1998sa,Zhu:1998am}. Therefore, the vale of the cosmological constant is greatly controlled by extrinsic size of trapped baryonic matter on the brane which causes the cosmological constant to be small.
The cosmological constant and its energy density can be obtained as
\begin{eqnarray}\label{rho}
\begin{array}{cc}
\Lambda_0 L_{\text{Pl}}^{2}=2.56\times10^{-122},\\
\rho_{0\Lambda}=(4.80\times10^{-3}eV)^{4},
\end{array}
\end{eqnarray}
which is well consistent with the observation \cite{Ade:2013zuv,Barrow:2011zp}.

Note that $f_\pi^{-1}$ is approximately equal to the Compton wavelength of the pion. Hence, the energy density of the cosmological constant approximately
is \begin{eqnarray}\label{3}
\rho_{0\Lambda}\approx
\frac{3}{2\pi}\left(\frac{4\pi}{g^2}\right)^2G_0m_{\pi}^6.
\end{eqnarray}
Remarkably, equation (\ref{3}) is identical to the scaling law proposed by Zeldovich \cite{zel}.

\section{Conclusion}
\label{sum}
If the cosmological constant originates from the vacuum energy of particle physics, it suffers
from a serious problem of its energy scale relative to the dark energy density today \cite{1}.
A reasonable attitude towards this open problem is the hope that quantum gravity will explain  that the vacuum does not gravitate, $\Lambda_{\text{vac}}=0$ \cite{Ellis}.
Hence, one may be able to argue that cosmological constant, or dark energy, has  really a gravitational-geometrical origin and is not connected to the vacuum energy.
To solve the cosmological constant problem, following \cite{Jalalzadeh:2013wza}, we introduced a new gravitational model in which our spacetime is embedded  in the ambient space with seven extra dimensions. The use of Nash-Morse implicit function theorem, lets the embedded spacetime to have a thickness. As a application, we considered the FLRW cosmological model to be embedded in $11D$ ambient space. Consequently, the induced gravitational constant is a function of the extrinsic normal curvature of spacetime. The violation of strong equivalence principle implies the age of universe as a time scale of variation and therefore the variation of $G_N$ cannot be expected to be larger than $10^{-11}y^{-1}$. This suggest that the induced extrinsic term in field equations is  of the cosmological constant type.
 According to the equation (\ref{13}), the present value of cosmological constant can be described
in terms of $4D$ gravitational constant, the strong interaction pion-nucleon coupling parameter and the size of overlapping of the quark and antiquark wave functions in the pions. This theoretical  obtained value of  cosmological constant is consistent with the observations.
\\
\\
{\bf Acknowledgment}

The authors would like to thank the anonymous referees for their valuable comments and
suggestions to improve the quality of the paper.\\
\\

\end{document}